\documentclass{article} 
\usepackage{iclr2023_conference,times}
\usepackage{graphicx}
\usepackage{float}

\usepackage{amsmath,amsfonts,bm}

\def\eqref#1{equation~\ref{#1}}

\def\1{\bm{1}}

\DeclareMathAlphabet{\mathsfit}{\encodingdefault}{\sfdefault}{m}{sl}
\SetMathAlphabet{\mathsfit}{bold}{\encodingdefault}{\sfdefault}{bx}{n}

\usepackage{hyperref}
\usepackage{url}

\title{Emulating Radiation Transport on Cosmological Scales using a Denoising U-Net}

\author{Mosima P. Masipa\\
  Department of Physics and Astronomy\\
  University of the Western Cape\\
  Bellville,Cape Town, 7535, South Africa \\
  \texttt{mosima101masipa@gmail.com} \\
  \And
  Sultan Hassan \\
  Center for Computational Astrophysics, Flatiron Institute \\
  162 5th Ave, New York, NY 10010\\
  \texttt{shassan@flatironinstitute.org} \\
  \AND
  Mario Santos \\
  University of the Western Cape\\
  Bellville,Cape Town, 7535, South Africa\\
  \texttt{mariogrs@gmail.com} \\
  \And
  Gabriella Contardo\\
  International School for Advance  Studies (Sissa)\\
  Via Bonomea, 265, 34136 Trieste TS, Italy\\
  \texttt{gabriella.contardo@sissa.it} \\
   \And
  Kyunghyun Cho \\
  New York University\\
  New York, NY 10012 \\
  \texttt{kyunghyun.cho@nyu.edu} \\
}

\iclrfinalcopy 
\begin{document}
\maketitle
\begin{abstract}
Semi-numerical simulations are the leading candidates for evolving reionization on cosmological scales. These semi-numerical models are efficient in generating large-scale maps of the 21cm signal, but they are too slow to enable inference at the field level. We present different strategies to train a U-Net to accelerate these simulations. We derive the ionization field directly from the initial density field without using the ionizing sources' location, and hence emulating the radiative transfer process. We find that the U-Net achieves higher accuracy in reconstructing the ionization field if the input includes either white noise or a noisy version of the ionization map beside the density field during training. Our model reconstructs the power spectrum over all scales perfectly well. This work represents a step towards generating large-scale ionization maps with a minimal cost and hence enabling rapid parameter inference at the field level.
\end{abstract}

\section{Introduction}

The epoch of reionization (EoR) refers to the time in the history of the Universe during which the birth of the first sources of light reionized the Inter-Galactic Medium \citep[IGM, for review see][]{Zaurobi2010}. Studying this era provides deeper insights into the evolution and formation of the earliest populations of galaxies. There are several ongoing and planned large-scale experiments, such as the Hydrogen Epoch of Reionization Array~\citep[HERA,][]{DeBoer2009} and the Square Kilometre Array~\citep[SKA,][]{SKA2009}, that are devoted to detecting the morphology of EoR (ionized bubbles distribution). These next-generation of surveys will provide huge amounts of high dimensional data sets of large-scale ionization fields, and hence a new generation of simulations and tools are needed to extract most information out of these highly non-linear fields.

\begin{figure}[H]
    \centering
    \includegraphics[scale=0.45]{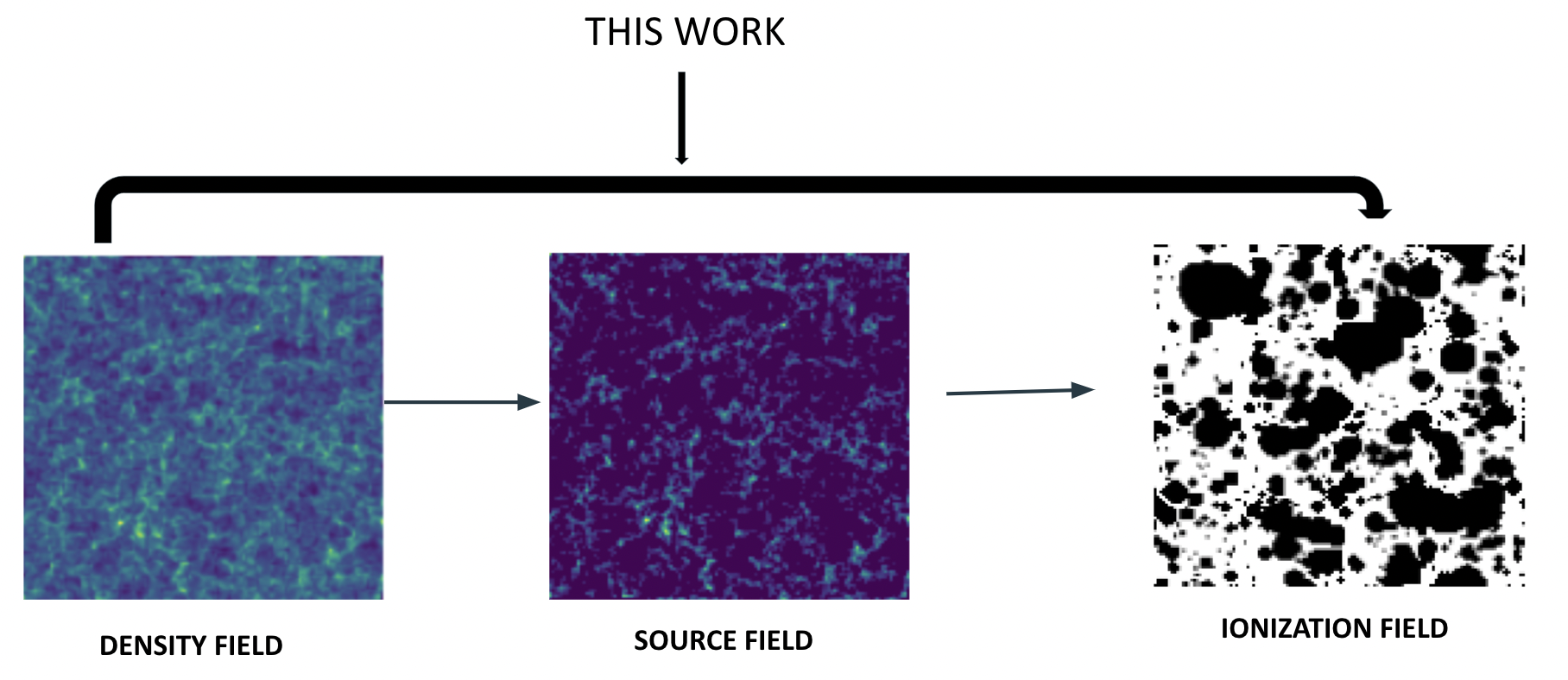}
    \caption{The basic workflow of all reionization simulations (e.g. SimFast21) is as follows: generating the initial density field (left), identifying sources (middle), and applying radiative transfer to obtain the ionized bubbles (right). Our goal is to skip the source identification and radiative transfer. We aim to directly convert the density field to ionization field using a U-Net}
    \label{fig:SIMFAST}
\end{figure}

Convolutional Neural Networks (CNNs) have been very successful to perform parameter recovery of model selection from ionization fields \citep[e.g.,][]{Hassan2019, Gillet2019, Mangena2020}. However, generating huge amount of data using current semi-numerical simulations of reionization, such as SimFast21\citep{Santos2010} or 21cmFAST\citep{Mesinger2010}, to fully explore the whole parameter space and to train CNNs is computational expensive. These semi-numerical models generate the ionization fields as follows: (i) An initial density field will be generated in the linear regime using a Gaussian distribution, which then will be evolved to the nonlinear regime by applying Zel'dovich approximation \citep{Zeldovich1970}. (ii) Sources (halos/galaxies ..etc.) will be identified using the well-known excursion set formalism \citep{Bond1991}. (iii) An approximated radiative transfer scheme will be applied to generate the ionization field out of the density and source fields. To accelerate these models, we aim to use different strategies to train a standard U-Net to skip the second and third steps (source identification and radiative transfer), which are both computationally very expensive. Figure~\ref{fig:SIMFAST} shows a random realization of the density, source, and ionization field. The top arrow indicates the aim of this work. This is a challenging task since we plan to generate a highly non-linear field (ionization) from a rather smoothed field (density) without feeding into the network the sources locations. 
\section{Method}
\begin{figure}[H]
    \centering
\includegraphics[scale=0.45]{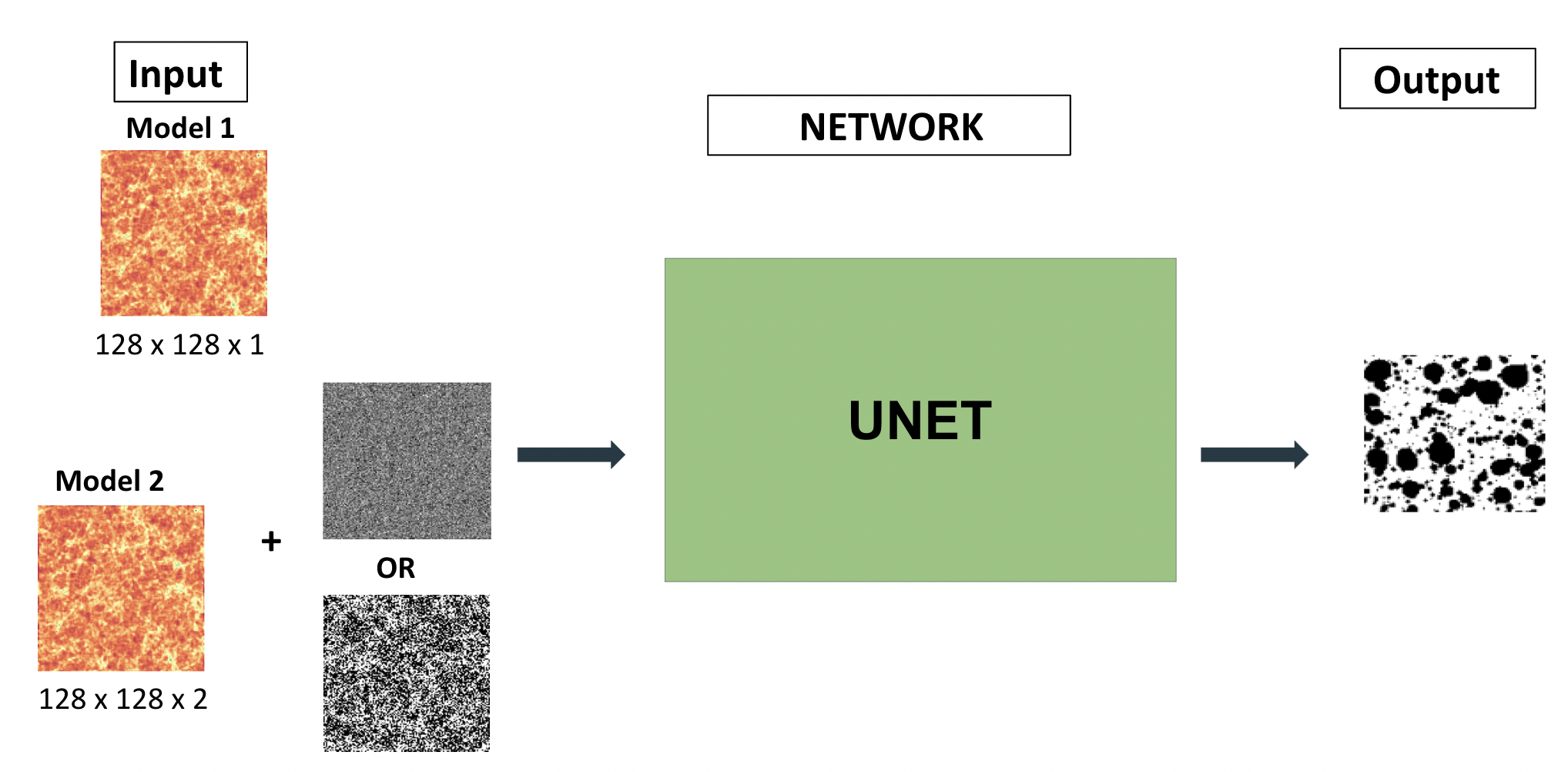}
    \caption{Visual summary of the different training strategies (Model 1 and 2) used in this study to train a standard U-Net.}
    \label{fig:UNet}
\end{figure}

\noindent To efficiently generate large-scale ionization fields, we closely follow a standard U-Net architecture, that was first used for image segmentation in the biomedical field ~\citep{Ron2015,Ibtehaz2019}. We have used the publicly available SimFast21\footnote{https://github.com/mariogrs/Simfast21} package to generate 15,600 reionization simulations with a box size of 250 Mpc on a side and a number of voxels of 128$^{3}$ to create a diverse training set of density and their corresponding ionization fields.
We split the data into an $80\%$  training set, $10\%$ validation set and $10\%$ testing set. As shown in Figure \ref{fig:UNet}, we use two different strategies to train our U-Net as follows:

{\bf $\bullet$ Model 1} takes a single input, namely the density field, to generate the ionization field. In this case, the method is deterministic and there exists a unique solution for each input. Hence, a single feedforward of the input through the network is sufficient for testing. 

\begin{figure}[H]
    \centering
    \includegraphics[scale=0.40]{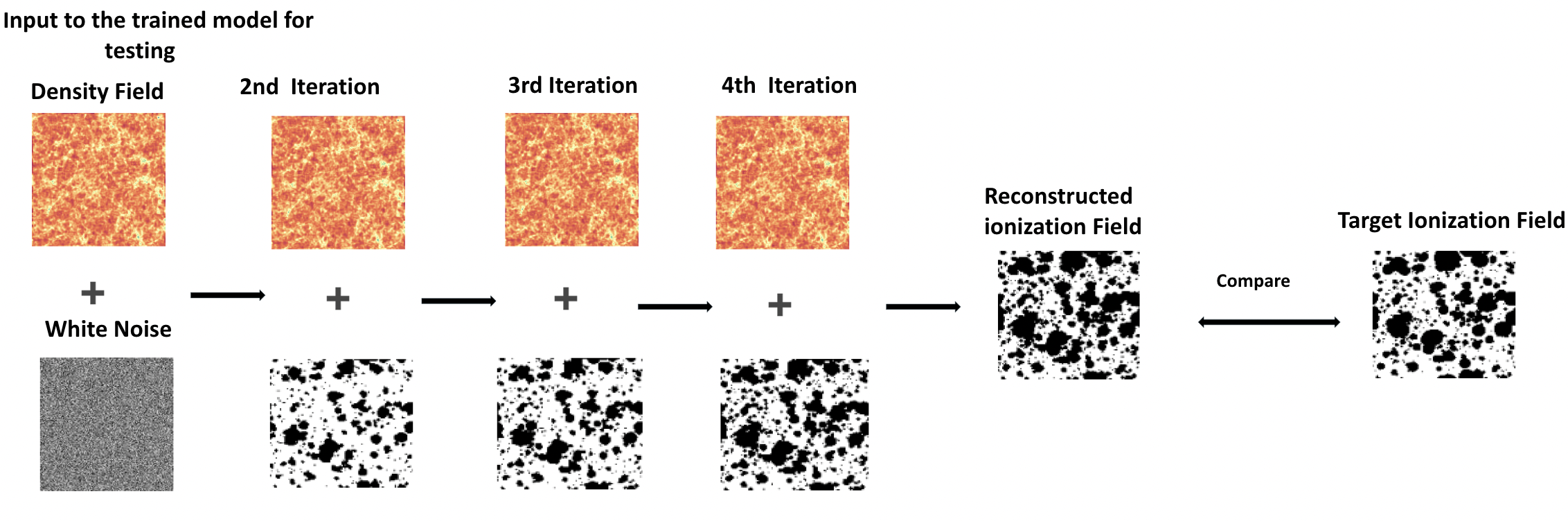}
    \caption{Visual summary of the recurrent testing adopted in Model 2 (the denoising U-Net). Four iterations are sufficient to produce a similar ionization field to the target.}
    \label{fig:testing}
\end{figure}

{\bf $\bullet$ Model 2} is a denoising U-Net that takes two inputs, namely the density field and either a white noise or a noisy version of ionization field during training, to generate the ionization field. In this case, the method is stochastic and there exists infinite solutions for each input, depending on the initial random seed. During testing, we first feedforward a white noise next to the density field. We then feedforward the output into the network to generate the second output. We continue to feedforward the output recursively till convergence in the loss evolution is achieved. In Figure~\ref{fig:testing}, we show the recurrent testing approach used in Model 2. The recursive testing is implemented using the following equation: $ y_n = \alpha y_{n-1} - (1-\alpha) $\rm U-net$ (x,y_{n-1}) $, where $x$
is the density field and $y_{n-1}$ is the previous prediction, and the regularization parameter $\alpha$ is set to 0.4 to prevent ionized bubble over growth.  We see that bubbles grow and the generated ionization map approaches the target gradually over iterations. We find that four iterations are sufficient to detect the bubble edges on different scales and achieve similar large scale power spectrum as will be seen later in Figure~\ref{fig:comparison}.

In all models, we use the same U-Net architecture, which consists of  5 convolutional layers to downsample and other 5 convolutional layers to upsample.  We use Adam as an optimizer, the Mean Squared Error (MSE) as the loss function, and the Rectified Linear Unit (ReLU) as the activation function. All models were trained using a single NVIDIA Tesla V100-32GB SXM2 GPU on Bridges-2, a high-performance computing platform from the Pittsburgh Supercomputing Cluster (PSC).  Model 1 was trained for 20 minutes and Model 2 for 3 hours to observe plateau in the loss evolution. Generating the ionization field takes only a fraction of a second as compared to 5 minutes using semi-numerical simulations, and hence a factor of at least 1,000 faster.

\section{Results}
\begin{figure}[H]
    \centering
    \includegraphics[scale=0.40]{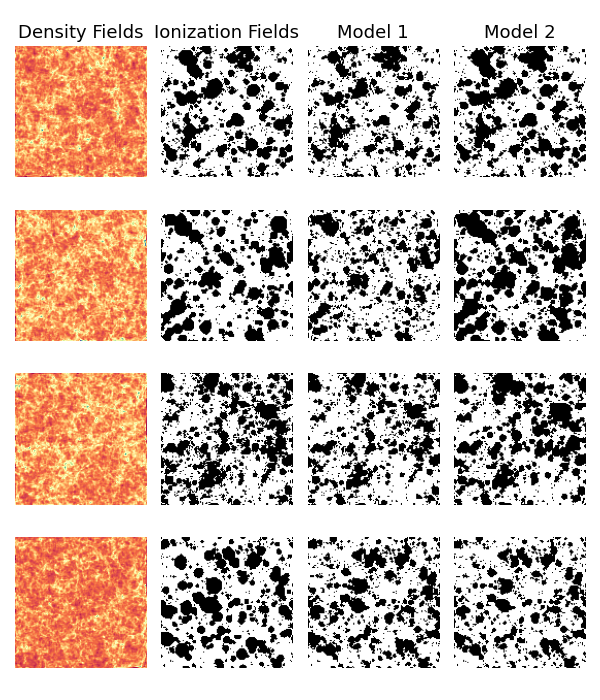}
    \includegraphics[scale=0.20]{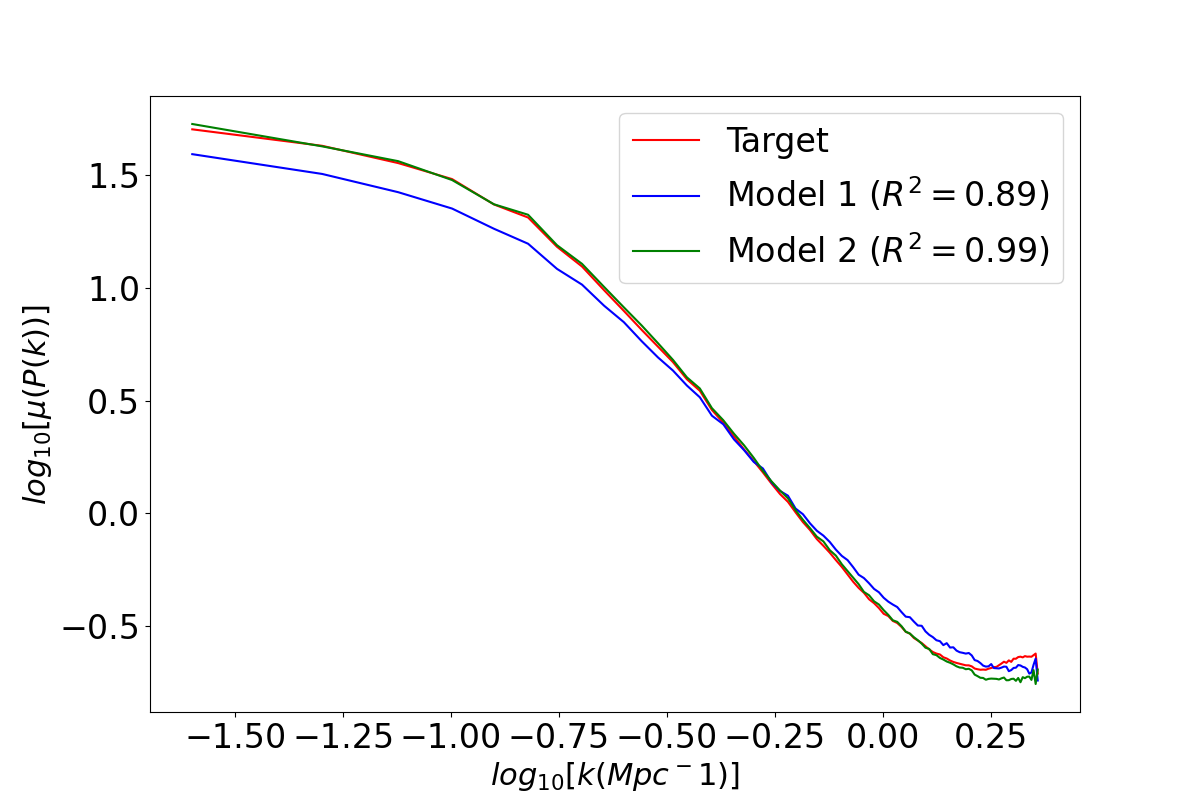}
    \caption{Left: Visualization of a random sample of images from the testing set. Right: The average power spectrum over all maps in the testing set. Model 2 is able to perfectly recover the target power spectrum over all scales.}
    \label{fig:comparison}
\end{figure}

Left panel of Figure \ref{fig:comparison} shows random realizations of the density field (first column), target ionization field (second column), generated ionization by Model 1 (third column) and by Model 2 (fourth column). The bubbles produced by Model 1 are smaller than those in the true ionization field, whereas Model 2 is able to produce similar large scale bubbles to the target. This effect can be clearly seen in the right panel of Figure \ref{fig:comparison}, where we show the average power spectrum ($\mu \,P)$ over all ionization maps in the testing set (red), as compared to those generated by Model 1 (blue) and by Model 2 (green). As quoted in the legend, Model 2 clearly outperforms Model 1, scoring a coefficient of determination $R^{2}$ of 0.99 as compared to 0.89. This is partly due to the fact that the network has seen the noisy version of the ionization fields (fields + white noise) during training. In addition, the recurrent testing implemented in Model 2 provides more flexibility to refine detection of bubble edges, leading to a more accurate reconstruction of the ionization field. 

\section{Conclusion}
This paper presents an attempt to emulate radiation transport on cosmological scales using a denoising U-Net. We have used two different strategies to train a U-Net to map out the ionization field directly from the initial density field without using the source locations.  Our best-performing model can predict the power spectrum perfectly well with $R^{2}=0.99$. As compared to semi-numerical models, our best-performing model is able to generate large-scale fields in a fraction of a second, leading to a factor of 1,000 faster. This work represents a step forward towards efficiently generating large scale ionization maps, and hence maximizing the scientific return of future reionization surveys. The training strategy presented in this work can be used in similar studies whenever the focus is to generate a highly non-linear signal out of a smoothed field.

Future works will include conditioning the ionized field generation on the amount of the neutral fraction and other astrophysical parameters such as the photon escape fraction to accurately emulate the behavior of simulations.



\label{Bibliography}
\bibliographystyle{unsrtnat}
\bibliography{iclr2023_conference}  


\end{document}